\documentclass[final,5p,times,twocolumn]{elsart_postprint}
\usepackage{hyperref}
\usepackage{cleveref}
\usepackage{verbatim}
\usepackage[OT1,T1]{fontenc} 
\usepackage[utf8]{luainputenc}
\usepackage{amsgen, gensymb, fdsymbol}
\usepackage{wasysym}
\usepackage{mathtools}
\usepackage{nicefrac}
\usepackage{doi}
\usepackage{tabularx}
\usepackage{ragged2e}
\usepackage{filecontents} 
\usepackage{graphicx}
\usepackage[doublespacing]{setspace}
\usepackage{verbatim}
\usepackage{float} 
\usepackage{booktabs}
\usepackage{etoolbox}
\usepackage{nameref}
\usepackage[export]{adjustbox}
\usepackage{xcite}
\usepackage{isotope,bpchem}
\usepackage{url}
\newcommand{\ra}[1]{\renewcommand{\arraystretch}{#1}} 
\providecommand{\e}[1]{\ensuremath{}\times 10^{#1}} 
\robustify{\textbf}
\journal{Scientific Reports \textbf{8}, 15926 (2018)}
\begin{document}
\begin{frontmatter}
\title{
Customisable X-ray fluorescence photodetector with submicron sensitivity\\ 
using a ring array of silicon p-i-n diodes
}
\author{Phil~S.~Yoon\fnref{psy}}\fntext[psy]{Email:~phil.s.yoon@gmail.com}
\address{P.O. Box 5000, Brookhaven National Laboratory, Upton, NY 11973-5000, USA}
\begin{abstract}
The research and development of silicon-based $X$-ray fluorescence detectors 
achieved its submicron sensitivity. Its initial use is intended for 
\emph{in-situ} beam monitoring at advanced light-source facilities. 
The effectively functioning prototype fully leveraged technologies and techniques 
from a wide array of scientific disciplines: $X$-ray fluorescence technique, 
photon scattering and spectroscopy, astronomical photometry, semiconductor physics, 
materials science, microelectronics, analytical and numerical modelling, 
and high-performance computing. 
In the design stage, the systematic two-track approach was taken 
with the aim of attaining its submicron sensitivity:
Firstly, the novel parametric method, devised for system-wide full optimisation, 
led to a considerable increase in detector's total solid angle (0.9 steradian), 
or integrated field-of-view ($\sim$ 3000 deg$^2$), thus in turn yielding 
a substantial enhancement of its photon-detection efficiency. 
Secondly, the minimisation of all types of limiting noise sources identified 
resulted in a boost to detector's signal-to-noise ratio, 
thereby achieving its targeted range of sensitivity.
The subsequent synchrotron-radiation experiment with this $X$-ray detector
demonstrated its capability to respond to 8-keV photon beams with 600-nanometre sensitivity.
This Article reports on the innovative and effective design methods
formulated for systematising the process of custom-building ultrasensitive photodetectors,
as well as future directions.
\end{abstract}
%
%
\end{frontmatter}  
%
\section{Introduction\label{sec:intro}}
\begin{figure*}[h!]
\centering\includegraphics[width=0.77\textwidth]{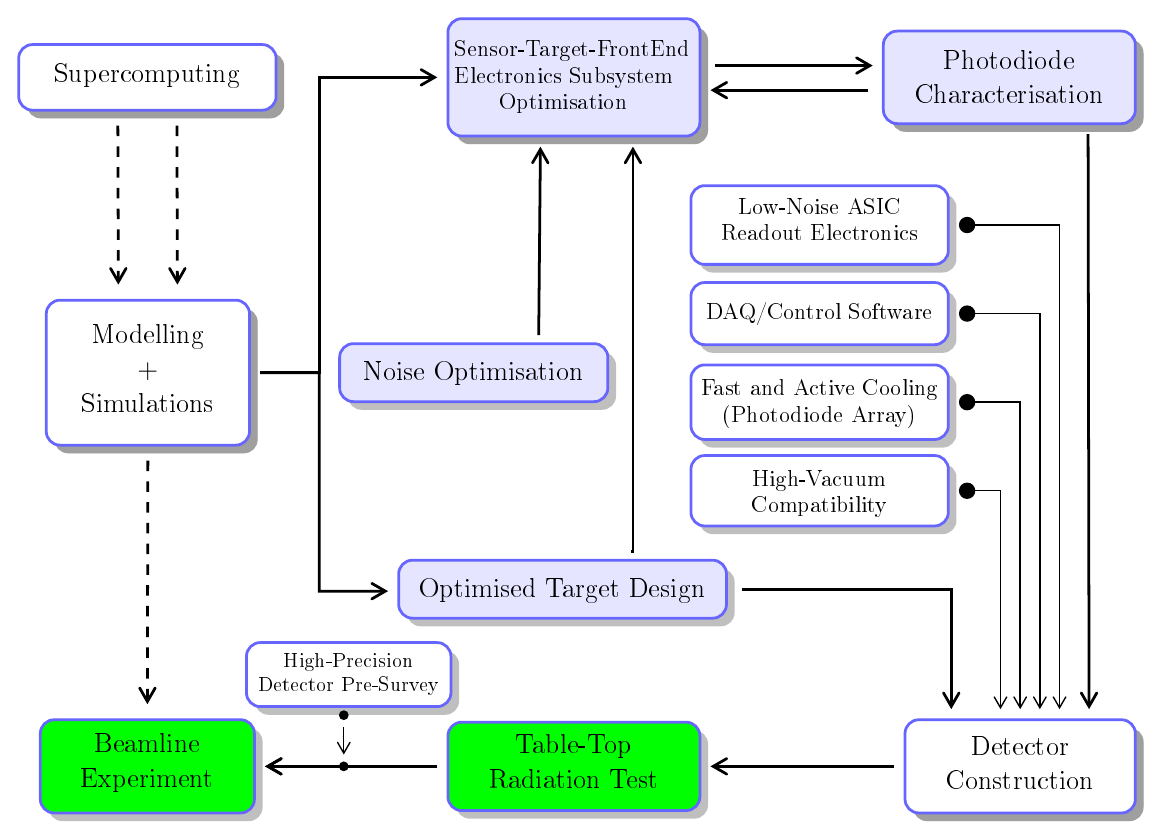}
\caption[Prototyping pipeline for the detector development.]{\textbf{Prototyping pipeline for the detector development.}
        ~Block diagram illustrating the workflow of developing an $X$-ray fluorescence detector with ultrahigh precision;
        shown in blue boxes is the four-phase design stage of optimising the core sensor-target-frontend subsystems.
        The calculation results from analytical modelling and supercomputer-aided 
        Monte-Carlo simulations provided quantitative and practical guidance (dashed line) 
        on conducting time-efficient beamline experiments.\label{fig:workflow}}
\end{figure*}

\begin{singlespace}
Contemporary synchrotron-based light-source facilities worldwide have been placing 
greater and greater user demand for \emph{in-situ} instruments having ultrahigh spatial sensitivity. 
The purpose of utilising such an ultrahigh-precision instrument is to fully realise the benefits 
of higher brightness and minuscule dimensions of $X$-ray beams radiating from a light source.
To this end, numerous groups\cite{alkire:quad_bpm, decker:pac2007, decker:biw2010, owen:pindiodes,
southworth:quad_pin_diode_xbpm, kenney:xbpm, plankett:pixelsensors, carg:indus2} have developed 
$X$-ray detectors of this type to fulfil their own needs over the past decades.
In the meantime, the advent of new synchrotron-radiation (SR) facilities
of this decade\cite{nsls2:jsr,nsls2:cdr} is sparking a strong need
to further increase photodetector's sensitivity to a scale of a fraction of one micron.
Driven by such growing needs for ultrahigh-precision beam-monitoring devices,
intensive R$\&$D efforts were dedicated to developing hard/tender $X$-ray detectors 
capable of monitoring nanometre-size photon beams {\itshape in situ}.
Apart from the pinpoint spatial sensitivity, 
the $X$-ray fluorescence (XRF)\cite{serpell:xrf,kinebuchi:xfa} technique 
was utilised, by design, for the semiconductor-based detector. 
Its initial applications are intended for downstream photon-intensive experiments. 
Using this photodetector of submicron sensitivity, 
$X$-ray beams focused onto beam-defining optical slits can be kept 
on the upstream side of minute biological samples under study 
(Fig.~\ref{fig:workflow}).
\par
At the inception of this R\&D project, a handful of design considerations 
were taken into account and thus led to building effectively functioning prototypes. 
The principal design goals pursued are listed in order of priority as follows:
(1) submicron-scale sensitivity,
(2) preservation of $X$-ray beam properties,
    such as photon flux and relative energy spread ($\Delta E/E$),
    on the downstream side of the detector,
(3) ultralow-noise operation,
(4) high-vacuum compatibility, and
(5) compact and cost-effective design.
As identified, these five considerations are foundational for
the design of position-sensitive detectors that are highly efficient 
for detecting photons\cite{gruner:xray_det}. 
The model-based full optimisations, 
based on which the detector prototype was constructed, 
gave the design a boost to its photon-detection efficiency 
and enabled a resulting enhancement of signals.  
In support of the low-noise design, active and fast detector-cooling 
modules and a dedicated control system for photon-counting application 
were integrated \textit{en masse} into the photon-detection system. 
Consequently, it was made possible to create a novel avenue towards 
a boost to detector's signal-to-noise (SNR) ratio and 
a pathway of achieving its ultrahigh spatial sensitivity.  
Additionally, utilising the parametric optimisation method can
empower detector designers to react nimbly to satisfy varying needs 
for individual beamline programs at light-source facilities.
Illustrated as a whole in Fig.~\ref{fig:workflow} is 
the codified process of prototyping a custom detector 
attaining the targeted sensitivity. 
\end{singlespace}
%
%
%
\section{Results\label{sec:results}}
\subsection*{Principle of operation\label{subsec:results}}
\begin{figure*}[h]
    \centering\includegraphics[max width=0.85\textwidth]{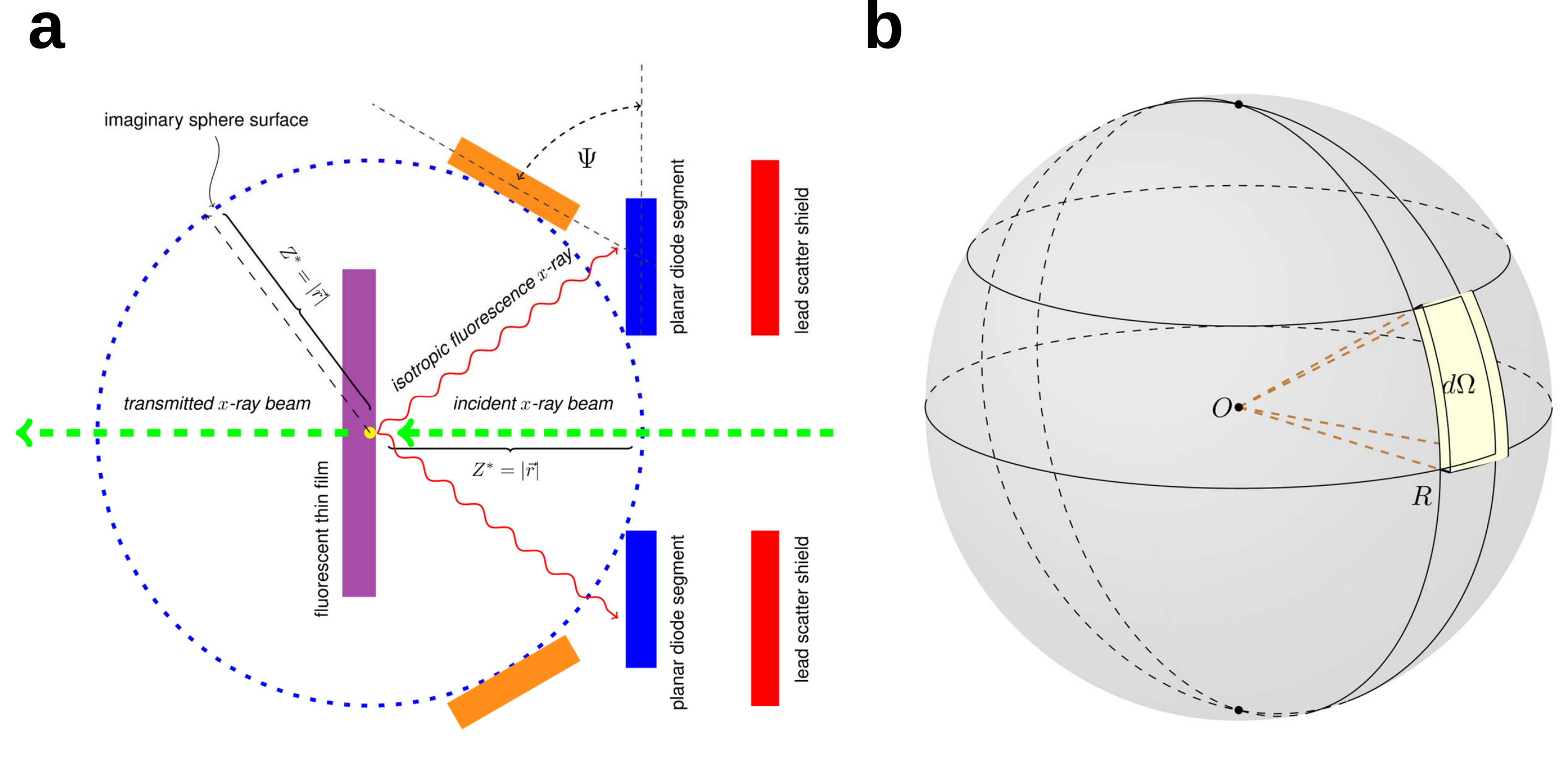}
    \caption[Principle of detector operation.]{\textbf{Principle of detector operation.}
    (\textbf{a})~Schematic presentation of the operational principle of the $X$-ray fluorescence detector.
                Drawn with the blue dashed line is the cutaway view of solid angle's imaginary sphere
                having its radius equal to the working distance $Z^{*}$.
                The projection angle $\Psi$ is formed between the sensor segment (solid blue) in the transverse plane 
                and the curvilinear sensor (solid orange) tangential to the spherical surface;
                note that the thin film, diode segments, and lead shield shavings are not to scale.
                (\textbf{b}) Solid angle, subtended at a secondary radiation source on the thin film,
                is defined on the surface of the imaginary sphere.\label{fig:backscattering}}
\end{figure*}

\begin{singlespace}
As illustrated in Fig.~\ref{fig:backscattering}(a),
the detector system design consists of four key functional components:
(1) a ring array of Si(PIN) diodes serving as a multisegmented photosensor,
(2) a thin metallized film as a source of fluorescence radiation,
(3) its dedicated ASIC-based readout electronics, and
(4) single-stage micro-thermoelectric (Peltier) coolers and a water-cooling module,
all of which are interfaced with a copper heat sink, or heat exchanger.
In essence, the sensor-target subsystems, surrounded by the readout electronics
and cooling modules, are situated at the very heart of the detector system.
Importantly, the thin fluorescing film--henceforth referred to as fluo-film, 
or fluo-target, or target for short--is introduced to the design 
in order to act as a source of secondary $X$-ray radiation emitted 
upon impingement of primary $X$-ray radiation.
As an addition, a lead (Pb) scatter shield, or visor,
is inserted immediately upstream from the sensor array,
so as to safeguard the ultrasensitive photosensors against
potential stray signals that may arise during operation.
And both upstream and downstream beryllium windows on a multi-port vacuum chamber 
are designed to be thin enough in consideration for downstream photon-starving experiments.
%
\par
As depicted in Fig.~\ref{fig:backscattering}(b),
the effective beam-through apertures are set to accept, at least,
a 5-$\sigma$ footprint of an incident beam. The dimension of this central aperture
accordingly limits the linear working range to a few millimetres
for a highly focused nanometre-size beam,
based on the 0.5-$\mu$rad tolerance for the beam-pointing stability required 
by the National Synchrotron Light Source II (NSLS-II) facility\cite{nsls2:cdr}.
Focused $X$-ray beams are then guided to propagate unimpeded
in a high vacuum, passing through the circular apertures
on both the sensor ring and the scatter shield.
Moreover, the fluorescing target is arranged in a configuration 
orthogonal to the propagating direction of incident $X$-ray radiation.
When primary radiation impinges upon the fluo-film,
a paucity of $X$-ray radiation\cite{serpell:xrf} emanates 
in the backward direction in the form of characteristic $X$-ray emission. 
As a result, the emission of nascent fluorescence radiation illuminates the backside of the sensor array.
At the same time, the vast majority of incident photons are transmitted.
In this process, the orthogonal beam-target topology becomes a critical factor
for monitoring a monochromatised beam with ultrahigh precision.
On this account, the orthogonal target configuration is required for the following three reasons:
First, the normal incidence forces an incident beam to experience from beam's vantage point
an uniform effective thickness of each fluo-target, regardless of its point of incidence.
Thus, the normal incidence of primary radiation on the target
makes it possible to preserve the incident beam properties during the process of
transmission and propagation towards a sample at the opposite end of the beamline from the light source.
Second, the on-target normal incidence, in turn, ensures isotropic illumination
and uniform photon detection over the entire active area of the planar sensor array.
And third, during transmission of $X$-ray beams through
the thin film, scientists can derive its beam centroid
from its entire beam profile, benefiting from uniform illumination
over the diode array.
On top of the orthogonal configuration, suppression of background-event signals is another critical issue.
Taken all together, the backward-scattering mode of operation was favoured over 
the forward-scattering mode in the light of minimising potential systematic uncertainties 
arising during operation.
The choice of the operation mode was borne out by the fact that
the former mode can suppress elastically scattered photons
(i.e., Rayleigh scatterings) coming from the target.
Consequently, the sensor array captures backward-scattered fluorescent radiation 
with higher detection efficiency and spectral purity.
Furthermore, three decisive advantages of using 
secondary $K_{\alpha}$-fluorescent radiation 
are the following\cite{perujo:k_alpha}:
(1) distinctively high intensity and variable photon flux,
(2) high spectral purity, and
(3) the availability of a wide selection of fluo-film materials
    with reference to the energy of incident radiation.
%
%
\begin{figure*}[h]
    \centering\includegraphics[width=0.87\textwidth]{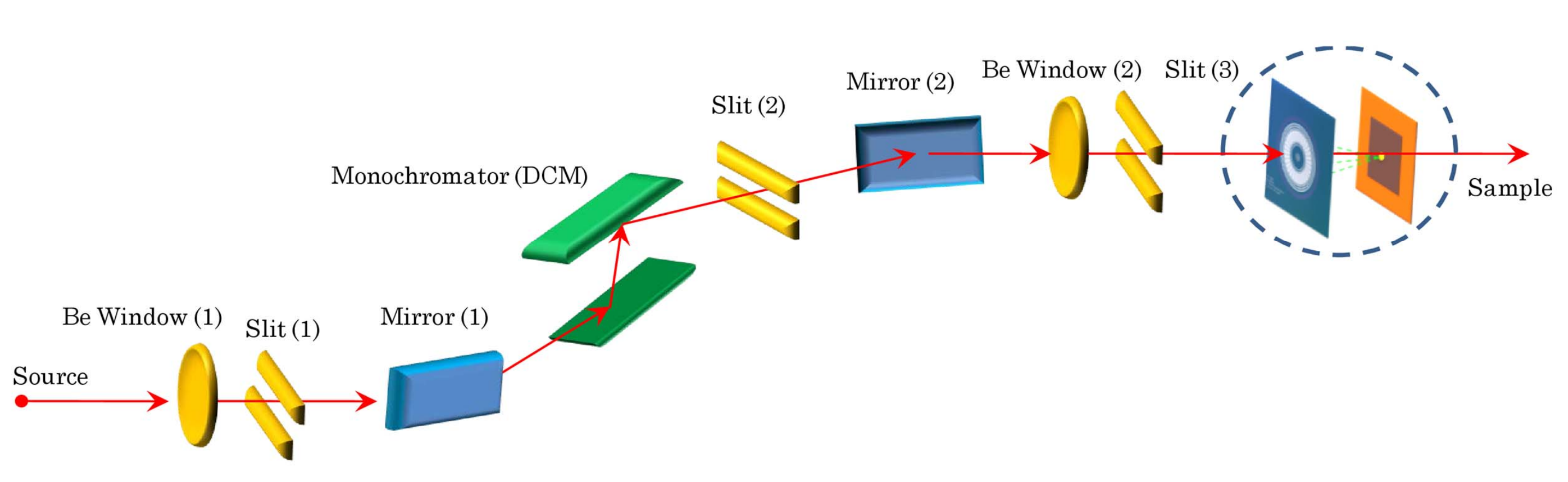} 
    \caption[Beamline layout.]{\textbf{Beamline layout.} An envisioned beamline layout is illustrated with the inclusion of
            the $X$-ray fluorescence detector assembly installed downstream from a string of various optical components
            (i.e., a double-crystal monochromator (DCM), mirrors, beam-shaping pre-slits, and an exit-slit collimator).
            One representative unit is shown to be positioned upstream from an experimental sample.\label{fig:beamline_layout}}
\end{figure*}
To visualise its full range of beamline operation, Fig.~\ref{fig:beamline_layout} shows
a schematic setup for monochromatic beamline components and the $X$-ray detector.
One of the design features of note is that its compact lateral form factor ($<$ 1 ft.)
makes it possible for each beamline to accommodate multiple colocated detectors.
This way a set of $X$-ray detectors operational along a beamlne enables 
extracting more beam parameters, such as divergence angles, beam emittances, and so forth.
\end{singlespace}
%
\subsection{Photosensor design\label{subsec:sensor}}
\begin{singlespace}
As an optical receiver, the photodiode ring array is coupled with 
a fluo-target and its dedicated front-end readout circuitry, 
all of which take the form of the trio core subsystems.
It is thus of foremost importance in designing a sensor ring array, 
that is most optimal for both the target and the readout electronics,
and a photodetector with low-noise signal processing.
%
\begin{figure*}[h]
    \centering\includegraphics[width=1.\textwidth]{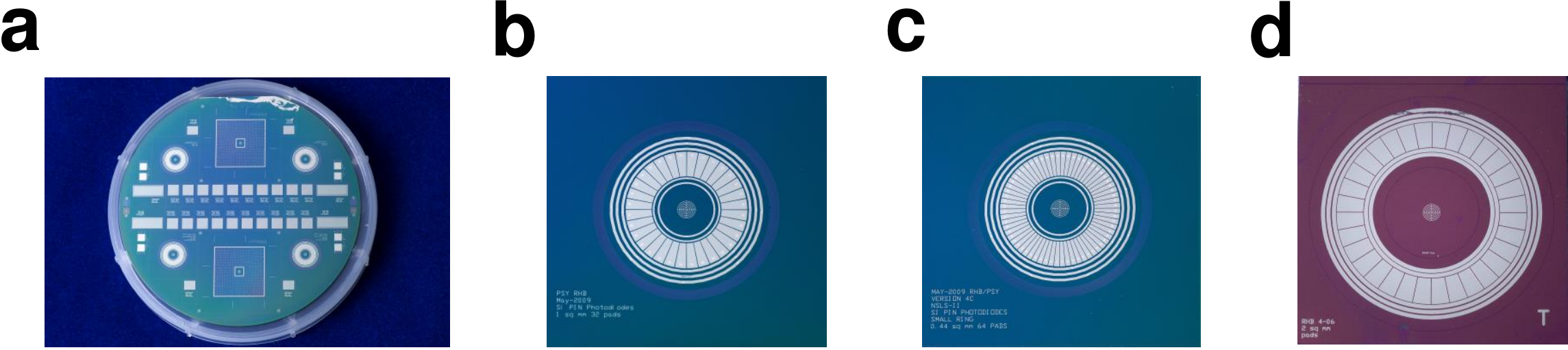}
    \caption[Prototype sensors.]{\textbf{Prototype sensors.} (\textbf{a}) 
             A finished $Si$ wafer containing four 32-segment diode rings arranged in each quadrant.
             (\textbf{b})~[Prototype-II-32] A fully optimised layout of the photodiode ring array
                           composed of 32 segments, each of which having its surface area of 1.0 $mm^{2}$.
             (\textbf{c})~[Prototype-II-64] A ring array of the 64 diode segments, each measuring 0.44 $mm^{2}$.
             (\textbf{d})~[Prototype-I-32] An unoptimised version of the 32-segment diode array.\label{fig:sensor}}
\end{figure*}
\par
Achieving the targeted sensitivity level presents two main challenges to the design:
(1) how to give rise to a substantial increase in photon-detection efficiency and
(2) how to realise ultralow-noise photon-detection and operation
under a high photon flux ($\gtr$ $10^{13}$~photons/s) environment.
The two-track approach was correspondingly taken to rise to the said challenges:
The first approach was to develop an effectual method of enhancing 
a higher degree of photon-detection efficiency, resulting in 
acquiring far more signals from the sensor array.
Unlike the prior developments elsewhere\cite{alkire:quad_bpm,
decker:pac2007,decker:biw2010, owen:pindiodes,southworth:quad_pin_diode_xbpm},
the sensor designs have the form of a multisegmented annular ring,
covering the full range of 2$\pi$-azimuthal angle, thus enabling highly efficient
capture of isotropic fluorescent radiation (Fig.~\ref{fig:sensor}(a)--(d)).
As the first step towards the photometric optimisation,
calculations of solid angle $\Omega_{ring}$\cite{bradt:astronomy}, 
subtended by the diode ring array at a source of fluorescence radiation, 
were worked out by virtue of finding real-valued analytical functions 
(Eqs.~(S1) and (S2)). 
Referring to Fig.~\ref{fig:backscattering}(b), 
the four-parameter solid angles, subtended at a point source, 
are expressed by Eqs.~\eqref{eqn:omega4a} and \eqref{eqn:omega4b}.
Here, $\Omega_{ring,4}$ and $\widetilde{\Omega}_{ring,4}$ denote 
respectively a solid angle for the planar sensor and a solid angle 
projected onto the spherical surface for the curvilinear sensor.  
\setlength{\abovedisplayskip}{15pt}\setlength{\belowdisplayskip}{15pt}
\begin{figure*}
\begin{subequations}
\begin{align}\centering
   &\Omega_{ring,4}
   =~2 N_{s}\Biggl\lbrace
       \arctan\Biggl( \frac{ \theta^{+}z^{*} }{ \sqrt{ (\rho^{-})^{2} 
       + (\rho^{-}\theta^{+})^{2} + (z^{*})^{2} } } \Biggr)
      - \arctan\Biggl( \frac{ \theta^{+}z^{*} }{ \sqrt{ (\rho^{+})^{2} 
      + (\rho^{+}\theta^{+})^{2} + (z^{*})^{2} } } \Biggr) \Biggr\rbrace \label{eqn:omega4a} \\
   &\widetilde{\Omega}_{ring,4}(z^{*};\theta^{+},\rho^{+},\rho^{-})
   =~\textup{Pr}(\psi) \cdot \Omega_{ring,4},\label{eqn:omega4b}
\end{align}
\end{subequations}
\end{figure*}
where the projection factor $\textup{Pr}(\psi)$ is defined as 
$\cos\psi$ (Fig.~\ref{fig:backscattering}(a)).
The variables $z^{*}$ and $N_{s}$ represent respectively
the working distance of the target from the sensor plane 
and the total number of diode segments in the array.
And $\theta^{+}$, $\rho^{+}$, and $\rho^{-}$ signify polar angle,
radial coordinate components of the sidelines of each diode segment, 
respectively (Supplementary Fig.~1).
In the case of a planar sensor (Fig.~\ref{fig:backscattering}(a)),
the projection factor $\textup{Pr}(\psi)$ is put into play 
for calculating solid angles with precision.
First and foremost, the application of this parametric method 
allows for quantitatively determining a \textit{practical} range 
of working distance optimal for the target in use.
As observed in Fig.~\ref{fig:ivcv}(a), the optimised sensor array
attains its total peak solid angle of 0.90 steradian, or
integrated field of view of $\sim$ 3000 $deg^{2}$,
under irradiation with photon beams having a 90-$\mu$m diameter
at an optimal working distance of 4.5 mm--the number of photon counts 
peaks out when target's working distance is at this calculated optimum position.  
Given these parameters, the photon-detection efficiency $\eta_{\small det}$
of the optimised core subsystems is estimated to be
$\eta_{\small det} = \Omega_{ring}/4\pi = \sum^{N_{d}}_{id=1}\Omega_{id}/4\pi = 0.072$.
Ultimately, this parametric method provides a quantitative basis for
determining the optimal layout and dimensions of the sensor array with 
the inclusion of multiple guard rings and the central aperture.
The foregoing analytical model made it possible to perform 
a series of rigorous analyses for fine-tuning the optimisation 
of the core subsystem.
According to the calculation results, two optimised versions
were fabricated---dubbed Prototype-II-32 and Prototype-II-64---
in reference to the unoptimised Prototype-I.
As Fig.~\ref{fig:sensor}(b)--(d) shows, 
the total surface areas of Prototype-II were halved,
whereas its peak solid angle and photon-detection efficiency 
doubled as a direct result of the photometric optimisation.
In addition, the full optimisation, by including spatial constraints 
imposed by indirect-bandgap semiconductors, produced a compact design, 
thus creating enough room for multiple concentric-circular guard rings.  
Accordingly, a system of two inner and three outer guard rings 
was incorporated in order to smoothly step down the electric potential 
to the ground (Fig.~\ref{fig:sensor}(b) and (c)).
As a side benefit, the resulting miniaturisation made its way
for lower-cost and higher-volume manufacturability on a single wafer.
And the second approach was the minimisations of the system-wide noise levels achieved in various ways.
As part of the optimisation process, the idea of multisegmentation was introduced
to the design of ring-shaped diodes. It was based on the fact that the multisegmented design
offers a few benefits to recognise: 
In the first place, the multisegmentation allows for the reduction in junction capacitance, 
thereby suppressing series noise and enabling sensor's faster response. 
In the second place, this multisegmentation provides the benefit of creating 
fine-tuning knobs for enabling capacitive matching between each diode segment 
and the front-end readout circuit. 
%
Upon sensor fabrication, full electrical characterisation was performed on
individual diode segments. Follow-up extraction of the characteristic
parameters in detail rendered quantification of detector's performance,
which was fed into the optimisation loop, as illustrated in Fig.~\ref{fig:workflow}.
%
\begin{figure*}[h!]
    \centering\includegraphics[width=0.9\textwidth]{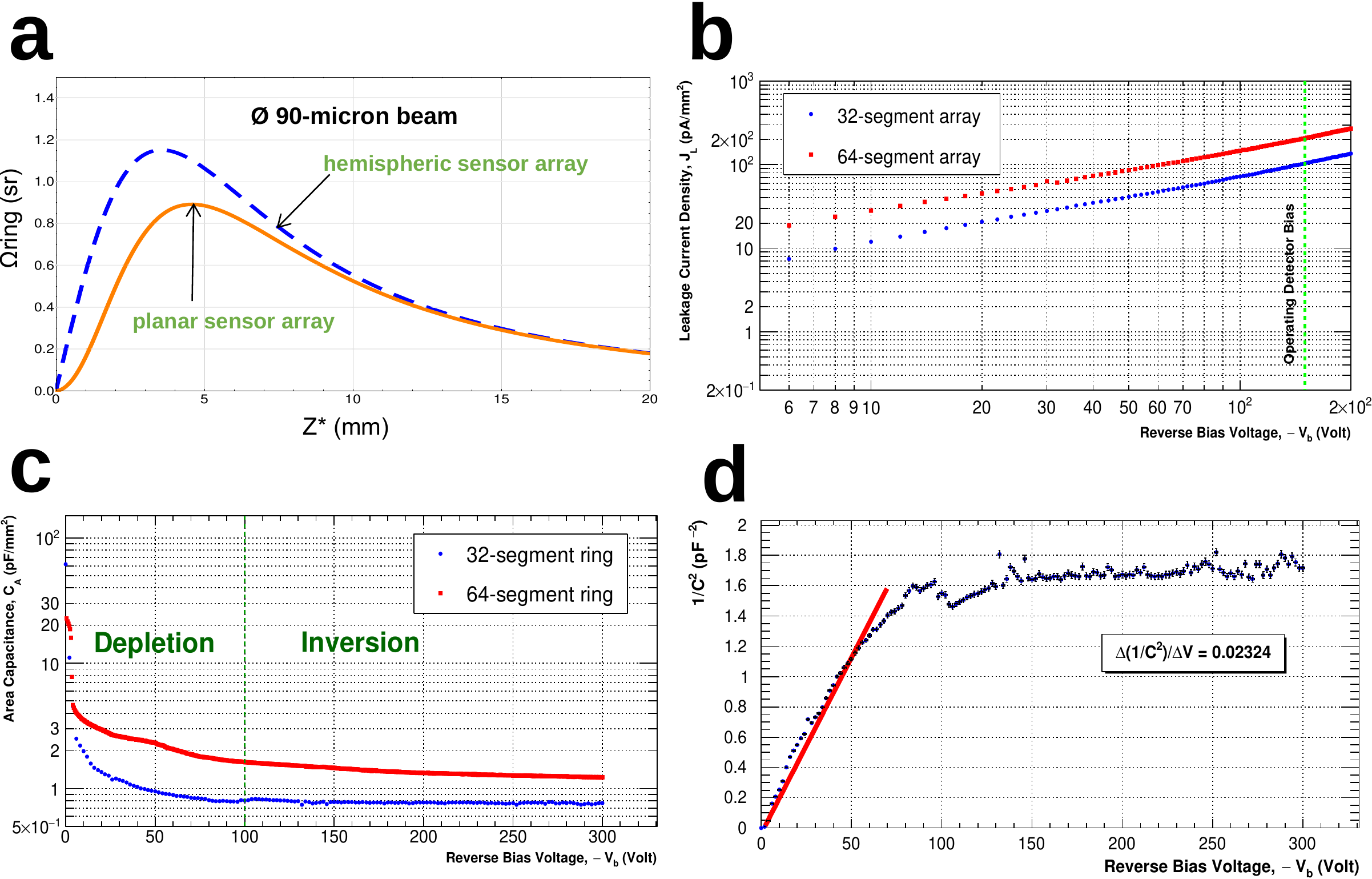}
    \caption[sensor characterization]{\textbf{Sensor characterisation.}~
              (\textbf{a})~[Prototype-II-32] Profiles of the total solid angles $\Omega_{ring}$ over working distance $Z^{\star}$.
                          The lower trace (red solid) corresponds to the optimised planar sensor array $\Omega_{ring}$, and 
                          the upper trace (blue dashed) to the optimised hemispheric array of sensors $\Omega_{hemi}$.
              (\textbf{b})~[Prototype-II-32/-64] 
              Measured $J_{r}$--$V_{r}$~characteristics of the photodiode segments at $\sim$ 300 K and 1 atm.;
                          the blue trace corresponds to the 32-segment ring and
                          the red trace to the 64-segment ring. Under the reverse-biased DC
                          voltage across the device up to 200 V, leakage current densities $J_{r}$
                          of the 32-segment and 64-segment ring array increase at $\sim$ 300 K.
                          The vertical green dashed line marks the operating detector bias of -150 V.
              (\textbf{c})~[Prototype-II-32] 
              Measured $C_{A}$--$V_{r}$ characteristics of the Si(PIN) diode segment at $\sim$ 300 K and 1 atm.;
                          plotted against the reverse DC bias on a log scale is the areal capacitance
                          undergoing the depletion and the inversion processes. 
                          The upper trace in red corresponds to the 64-segment array and 
                          the lower trace in blue to the 32-segment array. 
              (\textbf{d})~[Prototype-II-32]~Plot of $1/C_{A}^{2}$ as a function of reverse-biased DC sweep $-V_{r}$
                          showing the dopant concentration of the semiconductor junction device.
                          In the depletion region, a linear fit (red) is applied to the data points 
                          with systematic errors propagated.\label{fig:ivcv}}
\end{figure*}
When irradiated with photon beams of 90-$\mu m$ diameter,
the total solid angle of the planar sensor array peaks at 4.5 mm,
whereas an optimised hemispheric sensor array is expected to bring about 
a further increase of 30 $\%$ in its solid angle (Fig.~\ref{fig:ivcv}(a)).
It was observed from Fig.~\ref{fig:ivcv}(b) that surface leakage current 
density $J_{r}$ increases linearly with the reverse-biased voltage applied 
across the p-i-n junction with ohmic contacts.
In particular, the $C_{A}$--$V_{r}$ measurements took into account
leakage currents as a dissipation factor (Fig.~\ref{fig:ivcv}(c)).
Furthermore, dopant concentration was obtained from the slope of
the $1/C_{A}^{2}$--$V_{r}$ curve in the depletion region (Fig.~\ref{fig:ivcv}(d)).
And it was ascertained from Fig.~\ref{fig:ivcv}(b) that
the level of leakage current density $J_{r}$ is held below 100 $pA/mm^{2}$
under reverse-biased DC sweep down to $-150$ $V$ at room temperature,
while satisfying a set of the design criteria.
On the other hand, under the applied electric field, sensor capacitance decreases
until full depletion is reached at around $-150$ $V$ (Fig.~\ref{fig:ivcv}(c)).
In the inversion region, the areal capacitance $C_{A}$
as low as 0.5 pF was measured at the maximum depletion depth on 100 $kHz$,
based upon which the capacitive matching was conducted
(Fig.~\ref{fig:ivcv}(c))\cite{radeka:psd, poc:cmos}.
Listed as a summary in Table~\ref{tab:param} are salient detector parameters
extracted from the $J_{r}$--$V_{r}$ and $C_{A}$--$V_{r}$ measurements.
A silicon p-i-n (Si(PIN)) diode of high-speed response
produces sufficiently low leakage currents even at room temperature
without requiring a cryogenic cooling system.
Moreover, a wide dynamic range and excellent linearity are inherent to the silicon material.
Cognisant of these properties, silicon was opted over other semiconducting materials 
--e.g., Ge and Si(Li)--for the detection of hard/tender $X$-ray radiation (2 $\sim$ 25 keV).
Notably, the sensor array operates as a double-side junction--the frontside (device side) 
and the backside (window side)-illuminated structure\cite{rehak:sdd, carini:sdd}.
\begin{table*}[h]\centering\ra{1.2}
\caption{Figures of merit extracted for the array of 32-segment $Si$ diodes;
unless otherwise stated, each parameter was obtained from
$J_{r}$--$V_{r}$ and $C_{A}$--$V_{r}$ measurements taken at $\sim$ 300 K and 1 atm
with the reverse bias of 150 V and of an AC signal at 100 kHz.
Note that \emph{e} and \emph{h} denote electron and hole, respectively.\label{tab:param}}
\medskip
\renewcommand{\arraystretch}{1.} 
\begin{tabular}{l l}\toprule \hline\itemsep -7pt
        \textbf{Parameters}                          & \textbf{Values} \\ \hline
        total sensor thickness, $T$                  &  480 $\mu$m \\
        silicon oxide thickness, $T_{ox}$            &  560~nm \\
        depletion voltage, $V_{d}$                   & $\sim$ 150~$V$ \\
        depletion depth, $W_{d}$                     &  200~$\mu$m \\ 
        leakage current density, $J_{r}$             &  0.12 [260 K] / 104 [300 K]~$pA/mm^2$ \\ 
        effective dopant concentration, $N_{eff}$    &  5.0 $\e{12}$ $cm^{-3}$ \\
        resistivity, $\rho$                          & 1.0 $k\Omega$-$cm$ \\ 
        extrinsic Debye length, $L_{D}$              &  1.85 $\mu$m \\
        junction capacitance, $C_{j}$                & 0.5~$pF$ \\ 
        shunt resistance, $R_{sh}$                   & 180~$G\Omega$ \\
        carrier transit time, $t_{tr}$               & 10.2 [$e$]/34.1 [$h$] $ns$ \\
        rise time response, $t_{r}$                  & 1.9 $ns$  \\
        RC time constant, $\tau_{\small RC}$         & 0.9 $ns$ \\
        time resolution,  $\tau_r$                   & 2.8 $ns$  \\ 
        bandwidth, $f_{\small BW}$                   & 180~$MHz$ \\
        noise equivalent power ($\lambda$ = 229 nm)  & 1.6 $\e{-11}$ $W/\sqrt{Hz}$  \\
        specific detectivity, $D^{\star}$            & 6.4 $\e{10}$ $mm\sqrt{Hz}/W$ \\ \hline 
 \bottomrule
\end{tabular}
\end{table*}
\end{singlespace}
%
\subsection{Thin fluorescing film.\label{subsec:target}}
\begin{figure*}[h]
    \centering\includegraphics[max width=0.8\textwidth]{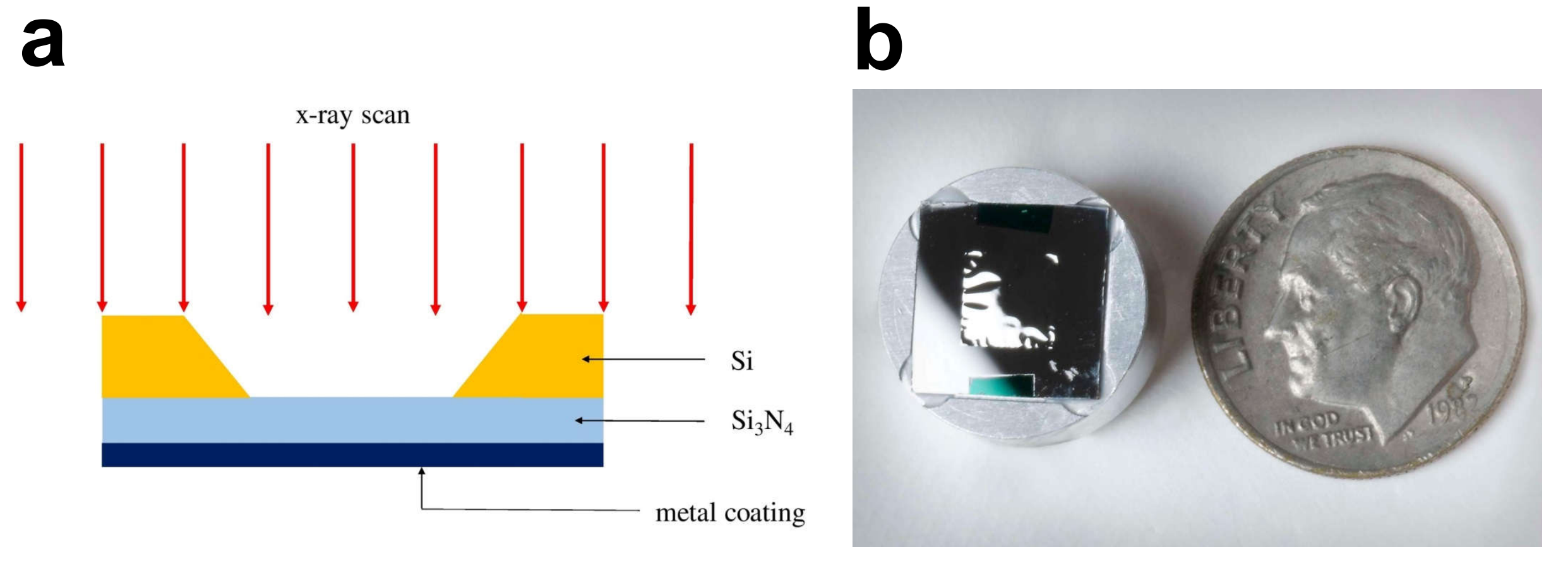}
    \caption[Fluorescing target.]{\textbf{Fluorescing target.}~\textbf{(a)}~Cutaway view of 
             the chromium (Cr) film vacuum-sputtered on the $Si_{3}N_{4}$ substrate window.
             \textbf{(b)}~Image of a fluorescing film mounted on an aluminium placeholder
             is shown with a US dime coin for comparison of dimensions.\label{fig:target}}
\end{figure*}

\begin{singlespace}
A high-transmission fluo-film optimal for the incident beam energy 
was constructed with the identification of the following elements to consider:
(1) the range of the energy of the $X$-ray beam in operation,
(2) the optimal range of beam energy to which the detector responds, 
(3) K-shell absorption edge of the fluorescing target material, and 
(4) the intensity expected in the $X$-ray beam.
A priori information described above sets standards for selecting appropriate film materials.
Hence, the selection of materials becomes a necessity, based on their K-shell values 
having sufficient separation below the energy of incident $X$-ray beams. 
The following is a list of the selection along with respective values of $K_{\alpha}$ and wavelength $\lambda$:
\BPChem{\_{22}}Ti ($K_{\alpha}$ = 4.511 keV; $\lambda$ = 2.749 $\mathring{A}$),
\BPChem{\_{24}}Cr ($K_{\alpha}$ = 5.415 keV; $\lambda$ = 2.290 $\mathring{A}$),
\BPChem{\_{25}}Mn ($K_{\alpha}$ = 5.899 keV; $\lambda$ = 2.102 $\mathring{A}$),
\BPChem{\_{26}}Fe ($K_{\alpha}$ = 6.404 keV; $\lambda$ = 1.936 $\mathring{A}$),
\BPChem{\_{27}}Co ($K_{\alpha}$ = 6.930 keV; $\lambda$ = 1.789 $\mathring{A}$).
The $X$-ray transmission efficiency and fluorescence signals contend with each other.
The optimal thickness of each of the film materials was determined from the existing
experimental data of mass-absorption coefficients $\frac{\mu}{\rho}$ from the literature
\cite{hubbell:survey, hubbell:photon_mass_atten_coeff, alkire:mass_absorp_coeff}
and numerical simulations that follow (Supplementary Fig.~2). 
In such a way, transmission rates of incident photon energy of 8-keV are above 90 $\%$
with a selection of differing metallic media for use as a thin film.
The transmission efficiency of an incident $x$-ray beam is important particularly
for photon-starving experiments being conducted at synchrotron light sources.  
The thickness of thin metal film is often too feeble to enable a self-supporting 
fluo-target of adequate quality. As a solution, a robust $X$-ray transmission window 
was introduced to the fluo-film design.
The substrate material of choice is the commercial silicon nitride ($Si_{3}N_{4}$) window 
(Norcada, Inc., Edmonton, AB, Canada: NX10500E, Supplementary Fig.~3),
which is known to be a low-stress ($<$ 250 MPa) radiation-hard substrate 
for $X$-ray applications\cite{gw:si3n4, toermae}.
For providing sufficient mechanical strength,
a 500-nm thick single-layered film was vacuum-sputtered
through a mask onto the silicon-nitride window
(5 mm $\times$ 5 mm $\times$ 500 nm)
at the NSLS-II in-house facility\cite{conley:multilayer}.
Displayed in Fig.~\ref{fig:target} (a) and (b) are the structure of the $X$-ray window 
and a thin film supported by the window substrate mounted onto an aluminium placeholder.
%
\end{singlespace}
%
\subsection{Photon-transport Monte-Carlo simulations\label{subsec:sim}}
\begin{singlespace}
During detector operation, background signals from coherent elastic
and incoherent inelastic scattering may be emitted anisotropically
in addition to fluorescent signals, owing to the polarisation of synchrotron radiation
in the plane of synchrotron\cite{landau:synchrad, schwinger:synchrad}.
For the quantitative assessment of signals and backgrounds,
Monte-Carlo (MC) simulation techniques were employed to study 
$X$-ray absorption, scattering, and fluorescent radiation.
Relying on the use of EGS4-based R\"Ontgen SImulator (R\"{O}SI version 0.19)
package\cite{rosi, namito:egs4}, the MC simulations were conducted with
the inclusion of the geometry of the entire detector assembly and the vacuum chamber.
Detector simulations, using realistic statistics of photon flux ($\sim$ $10^{12}$~photons/s),
were carried out on high-performance supercomputing platforms\cite{nersc}.
The primary purposes of such high-fidelity MC simulations are the following:
(1) quantification analysis of detector performance and
(2) estimation of potential background events coming from ambient scatterings,
    both of which are important for attaining its submicron-scale sensitivity.
This MC model collectively included signals from $X$-ray fluorescence and 
Rayleigh and Compton scatterings. In particular, the angular distribution 
of Rayleigh photons of a few keV, which are scattered from medium- or high-Z materials, 
is confined mostly in a wide-angle cone open in the forward direction\cite{klein_nishina, nelms_oppenheim}.
For the photon energy ranging from 10 to 50 keV, the photoelectric interaction
is the dominant signal process, having sole responsibility for the fluorescence phenomenon.
On the other hand, Rayleigh scattering is the dominant background process
over Compton scattering in the energy range of 5 $\sim$ 25 keV.
With the selection of materials 
(i.e., \BPChem{\_{24}}Cr, \BPChem{\_{25}}Mn, \BPChem{\_{26}}Fe, and \BPChem{\_{27}}Co),
MC studies show that estimated contributions from the two competing interactions
--Rayleigh and Compton scatters--are as insubstantial as below
the level of 1.0 $\%$ and 0.1 $\%$, respectively. 
As defined in equation~(\ref{eqn:purity}),
the photon purity $P_{K,fluo}$ of the K-shell fluorescent radiation
is the ratio of K-shell fluorescent radiation $N_{K,fluo}$
to a sum of scattered radiation $N_{scatt}$ and $N_{K,fluo}$:\vspace{-1ex}
\begin{equation}\label{eqn:purity}\centering
P_{K,fluo} = N_{K,fluo}/(N_{K,fluo} + N_{scatt})
\end{equation}
Further, the MC simulations indicated that the purity $P_{K,fluo}$
for each individual element in the selection
amounts to nearly 100 $\%$ in a vacuum.
\end{singlespace}
%
\subsection{Readout electronics and peripheries\label{subsec:readout}}
%
\begin{figure*}[h!]
\centering\includegraphics[width=1.\textwidth]{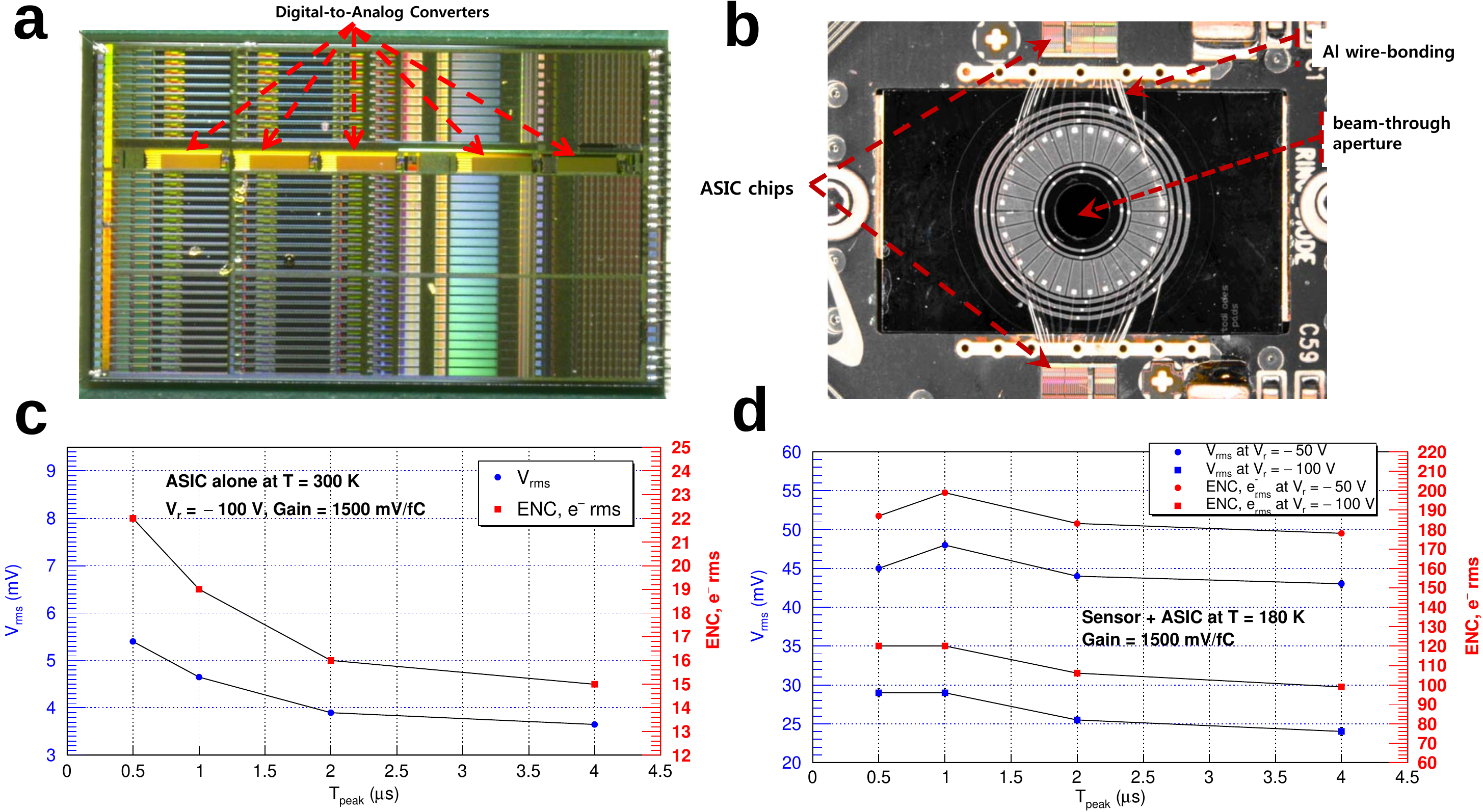}
\caption[Sensor-ASIC interconnections and front-end readout noise performance.]{
\textbf{Sensor-ASIC interconnections and front-end readout noise performance.}
(\textbf{a})~Image of the High-Energy Resolution Multi-Element Spectrometer (HERMES4) ASIC chip
of which layout dimension is 3711 $\mu$m $\times$ 6287 $\mu$m.
(\textbf{b})~Details of the sensor-ASIC layout displaying on-chip interconnections. 
(\textbf{c})~Analysed was noise originated from the HERMES4 ASIC chip alone 
without photogenerated current at 300 K.
The blue circles in the plot represent rms voltages in units of mV,
while the red squares do the measured equivalent noise charge ($e^{-}$ rms)
at different selections of peaking time $T_{peak}$
(0.5, 1, 2, and 4 $\mu$s). Note that the gain is set to 1500 mV/fC.
(\textbf{d})~Plotted as a function of $T_{peak}$ are the rms voltages
arising from the sensor array wire-bonded to the
two ASIC chips at near cryogenic temperature ($\sim$ 180 K).\label{fig:asic}}
\end{figure*}
\begin{singlespace}
Compact in-vacuum readout electronics were designed, utilising
dedicated application-specific integrated circuits (ASICs)
for the high-rate photon-counting application.
The HERMES4 ASIC, based upon 350-nm CMOS technology,
offers 32 channels of low-noise charge amplification,
high-order shaping with baseline stabilisation,
and peak detection for various low-noise analogue/digital processing
(Fig.~\ref{fig:asic}(a))\cite{degeronimo, cook:czt}.
For reducing stray capacitance and for suppressing signals stemming from electrical interconnections,
each sensor segment in the array is wire-bonded directly to the input of the front-end readout
channel (Fig.~\ref{fig:asic}(b)). As such, Al-wire wedge bonding and 
the ultrasonic technique were utilised between the input of an ASIC channel
and a bonding pad implemented on each diode segment.
According to the measurements, ASIC noise optimisation
enables an electronic resolution of 15 $e^{-}$ rms (equivalent noise charge, or ENC)
with a choice of 4-$\mu$s peaking time at room temperature.
With multisegmented photosensor's high finesse,
the event rate per each ASIC channel is diminished to 40 kHz 
in a high-flux SR condition. 
As a result, the front-end readout with 
low-power consumption (8 mW/channel) was realised.
For high-flux applications, one major challenge is how to design
associated readout electronics capable of processing sufficiently
high photon statistics without reaching charge saturation.
It was observed during a beamline experiment that the processing rate
of this ASIC readout chip can cope with $\sim$ 100 kHz per channel. 
Additionally, the integrated cooling subsystem functions to lower the operating temperature
of the Si(PIN)-diode array nestled on the readout printed circuit board.
Hence, the active and fast detector cooling is an added functional feature
that helps minimise persistent system-wide parallel noise.
In this respect, the detector system is outfitted with efficient cooling modules in three ways:
(1) A ring array of thermoelectrically-cooled diodes is included 
in the power budget for minimising bulk leakage currents.
(2) Both the readout electronics and the thin film
make exceptionally good thermal contact with a water-cooled heat sink copper block.
And, (3) applying adhesive with high thermal conductivity
(5.77 W/m\,K) for interconnecting individual components yields
a remarkable enhancement of the heat-transfer process.
As a consequence, its operating temperature can plummet to
$-40~\celsius$ in as fast as a few seconds after power-up.
As an integral component of the detector system,
the mechanical support made of an Invar 36 (Alloy 36)\cite{patent:invar}
was designed together. 
This instrument support system of high thermal stability (100 nm/$\celsius$/hr)
ensures the steady maintenance of detector's high-precision alignment 
and spatial sensitivity in a temperature-controlled environment
throughout each year.
\end{singlespace}
%
\subsection{Radiation experimentation\label{subsec:expt}}\vspace{1ex}
%
\begin{figure*}[h!]
    \centering\includegraphics[width=1.0\textwidth]{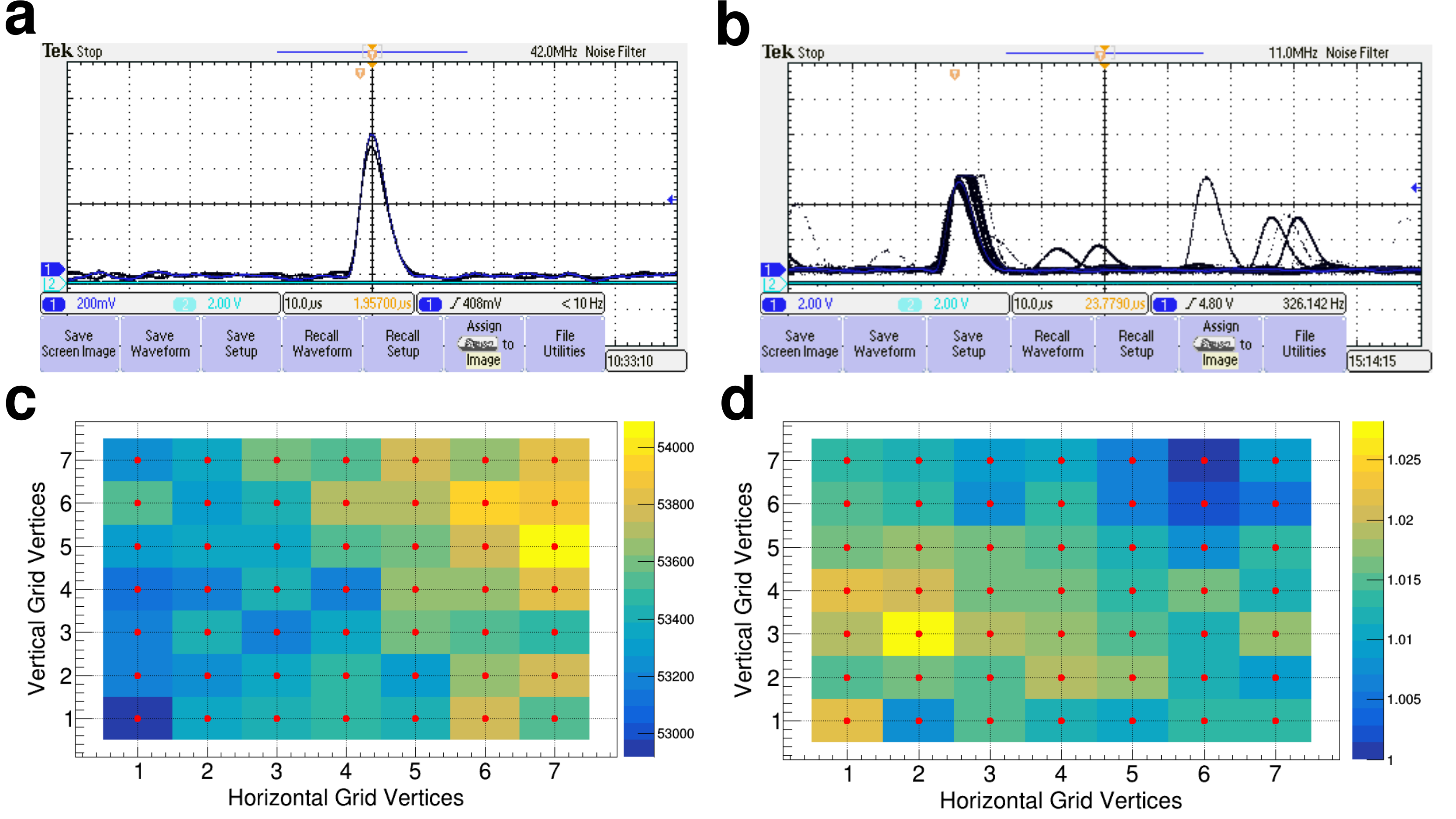}
    \caption[Beamline experimentation.]{\textbf{Beamline experimentation.}
             Real-time images of pulse-shaped output waveforms captured on oscilloscope (Tektronix MSO 4000 series) 
             display during detector operation at $-40~\celsius$,
             (\textbf{a})~when irradiated with 5.9-keV radiations from Fe-55 sealed sources 
                          in a high-vacuum condition at the base pressure of $10^{-4}$ torr, and
             (\textbf{b})~when irradiated by a synchrotron radiation in a high-vacuum condition of 
                          3 $\times$ $10^{-6}$~torr.
             (\textbf{c})~Mesh plot showing photon counts processed from the ASIC channel 34 
                          at each vertex (red dot) over the 7~$\times$~7 mesh grid. 
             (\textbf{d})~Photon counts processed from the channel 46 are normalised to the minimum count 
                          at each vertex and are then projected onto the same 7~$\times$~7 mesh grid.\vspace{2ex}\label{fig:blexpt}}
\end{figure*}
\begin{singlespace}
As a proof of concept, a table-top experiment was carried out first at room temperature 
with an Fe-55 sealed radioactive source, confirming the acquisition of noise-free
detector response (Fig.~\ref{fig:blexpt} (a)).
Next, the prototype detector, loaded with the fully optimised 32-segment sensor 
(Prototype-II-32) and target, was experimented with polychromatic beams 
at the X7A beamline of the National Synchrotron Light Source.
Prior to the SR experiment, high-precision alignment ($\sigma$ $<$ 60 $\mu m$) 
of the detector assembly was ensured at the beamline in an effort to 
minimise systematic uncertainties. 
The amount of detector noise, intrinsic to the radiation sensor and the front-end readout system,
was minimised consistently during both the table-top and the SR beamline experiment.
As evidenced by the experimental observations, the noise floor 
was brought down to the bottom level at $-40~\celsius$ in high-vacuum conditions 
(Fig.~\ref{fig:blexpt}(a) and (b)).
Then, a subsequent SR experiment was dedicated to one full round of 
beam-based transverse aperture scan with $X$-ray beams of 90 $\mu m$ 
in diameter in regular steps of 600 nanometres (nm) over 7~$\times$~7 mesh grid.
%
As cogent evidences, the variations of photon counts from each channel 
are mapped to the colour scales on the right-hand column and projected 
over the mesh grid (Fig.~\ref{fig:blexpt}(c) and (d)). 
This way the mesh plots visualise the variations of photon counts at 
each of the 49 vertices with the submicron interval.
Hence, these colour maps clearly delineate that the detector design 
has its capability to respond with 600-nm sensitivity and better under 
intense irradiation of $X$-ray photon beams at the given energy.
\end{singlespace}
%
%
\section{Discussion\label{sec:discussion}}
\begin{singlespace}
Leveraging the systematic design method, developed for fully optimising 
the detector system and for eliminating noise sources, maximised its SNR 
and thus boosted its spatial sensitivity.
Consequently, the long-standing barrier of detector's submicron sensitivity 
has been broken down. This breakthrough was achieved by amalgamating technologies 
from manifold areas of research fields.
Eventually, the custom-design method, offering great flexibility 
to add more new ideas and features, became a product of the transdisciplinary approach,
thereby ushering in the realm of nanometre-range sensitivity.
In a conventional prototyping process, 
a significant amount of lead time and cost 
is needed for completing an entire detector prototyping process.
By contrast, taking the two-prong approach reported in this Article
makes it possible to design an archetype of the photodetector 
reaching a desired level of sensitivity 
at a fraction of the previously required time and cost.
Above all, the detector design, ensuring normal incidence of primary radiation,
is apposite particularly to highly focused monochromatised $X$-ray beams,
expecting its surpassing photometric performance.
Further enhancements of its detection efficiency and sensitivity will
involve optimising the sensor-target subsystem based on a beam dimension 
and developing hemispheric, or semi-hermetic 2$\pi$-photodetectors
open in the forward direction. 
The calculations in Fig.~\ref{fig:ivcv}(a) also corroborate 
the concave photodetector systems designed elsewhere\cite{ko:eye_camera,someya:eyeball}.
Hence, one practicable solution to suggest is building 
such an optimised concave photodetector in a honeycombed structure.
The benefits from utilising such ultrahigh-precision instruments 
are expected to ripple out across the scientific community. 
Further, the position-sensitive detector of fluorescence type, 
realised by the tailored design process, has a likely impact on 
applications of wide and far-reaching appeal to 
the scientific and industrial research alike.
\end{singlespace}
%
\bibliographystyle{unsrt}
\bibliography{xdetsrep}
\section{Acknowledgements}
This R\&D work was conducted under the aegis
of the U.S. Department of Energy,
Office of Basic Energy Sciences,
by Brookhaven National Laboratory (BNL)
under contract number DE-AC02-98CH10886.
The author acknowledges
the provisions and arrangements made
by the detector development group and
the beamline support group of
National Synchrotron Light Source-II,
and the Instrumentation Division of BNL throughout.
\end{document}